\documentclass[prl,aps,twocolumn,longbibliography]{revtex4-1}
\usepackage{color}
\usepackage{graphicx}
\usepackage{bm}
\usepackage{amsmath}
\usepackage{amssymb}
\usepackage{amsthm}
\usepackage{amsfonts}
\usepackage{enumerate}
\usepackage{latexsym}
\usepackage{float}

\usepackage[colorlinks=true,urlcolor=blue,citecolor=blue,allcolors=blue]{hyperref}
\newcommand{\bs}{\boldsymbol}

\usepackage{braket}
\usepackage{array}
\usepackage{float}
\usepackage{cleveref}

\renewcommand{\vec}{\mathbf}

\newcommand{\mc}{\mathcal}

\newcommand{\ceil}[1]{\lceil #1 \rceil}

\usepackage[dvipsnames]{xcolor}

\begin{document}

\title{Quantum Hall effect and Landau levels without spatial long-range correlations}
\author{Isac Sahlberg, Moein N.~Ivaki, Kim P\"oyh\"onen and Teemu Ojanen}
\affiliation{Computational Physics Laboratory, Physics Unit, Faculty of Engineering and
Natural Sciences, Tampere University, P.O. Box 692, FI-33014 Tampere, Finland}
\affiliation{Helsinki Institute of Physics P.O. Box 64, FI-00014, Helsinki, Finland}

\begin{abstract}
The spectrum of charged particles in translation-invariant systems in a magnetic field is characterized by the Landau levels, which play a fundamental role in the thermodynamic and transport properties of solids. The topological nature and the approximate degeneracy of the Landau levels are known to also survive on crystalline lattices with discrete translation symmetry when the magnetic flux through a primitive cell is small compared to the flux quantum. Here we show that the notion of Landau levels and the quantum Hall effect can be generalized to 2d non-crystalline lattices without spatial long-range order. Remarkably, even when the spatial correlations decay over microscopic distances, 2d systems can exhibit a number of well-resolved Landau-like bands. The existence of these bands imply that non-crystalline systems in magnetic fields can support the hallmark quantum effects which have been typically associated with crystalline solids.

\end{abstract}
\maketitle

\emph{Introduction--} There are few notions more fundamental and important to condensed-matter physics than Landau levels \cite{ezawa2013quantum}. A spectrum of charged particles in a magnetic field is squeezed into a sequence of flat bands $E_n$ that can support extensive degeneracy, giving rise to a rich variety of phenomena in solids \cite{girvin,solyom}. The classical examples include quantum oscillations in thermodynamic and transport quantities, such as the de Haas-van Alpen and the Shubnikov-de Haas effects, while the discovery of the quantum Hall effect ushered in the modern age of topological matter~\cite{PhysRevLett.45.494,PhysRevB.23.5632, PhysRevLett.49.405}. In gapless systems, the Landau levels are characterized by their linear dependence on the magnetic field $E_n\propto B$, and their behavior as a function of the level index $n$ which depends on the specific dispersion of the particles. The two most studied cases are the parabolic dispersion and the Dirac-type dispersion which correspond to $E_n\propto n$ and $E_n\propto \sqrt{n}$. The different $n$-dependence makes it explicit that the Landau level spectrum depends on the underlying band structure. This naturally raises the question of what happens to the Landau levels and the related phenomena in non-crystalline systems, where the notion of a band structure does not apply. More generally, how are the magnetic-field-related phenomena modified in non-crystalline systems? Although the effects of disorder on Landau levels have been studied in great detail~\cite{huckestein,PhysRevLett.76.1316,PhysRevLett.76.975,ando1974theory}, these studies do not apply to systems where a band structure cannot be assumed as a starting point for the analysis.
\begin{figure}[h!]
\includegraphics[width=0.93\columnwidth]{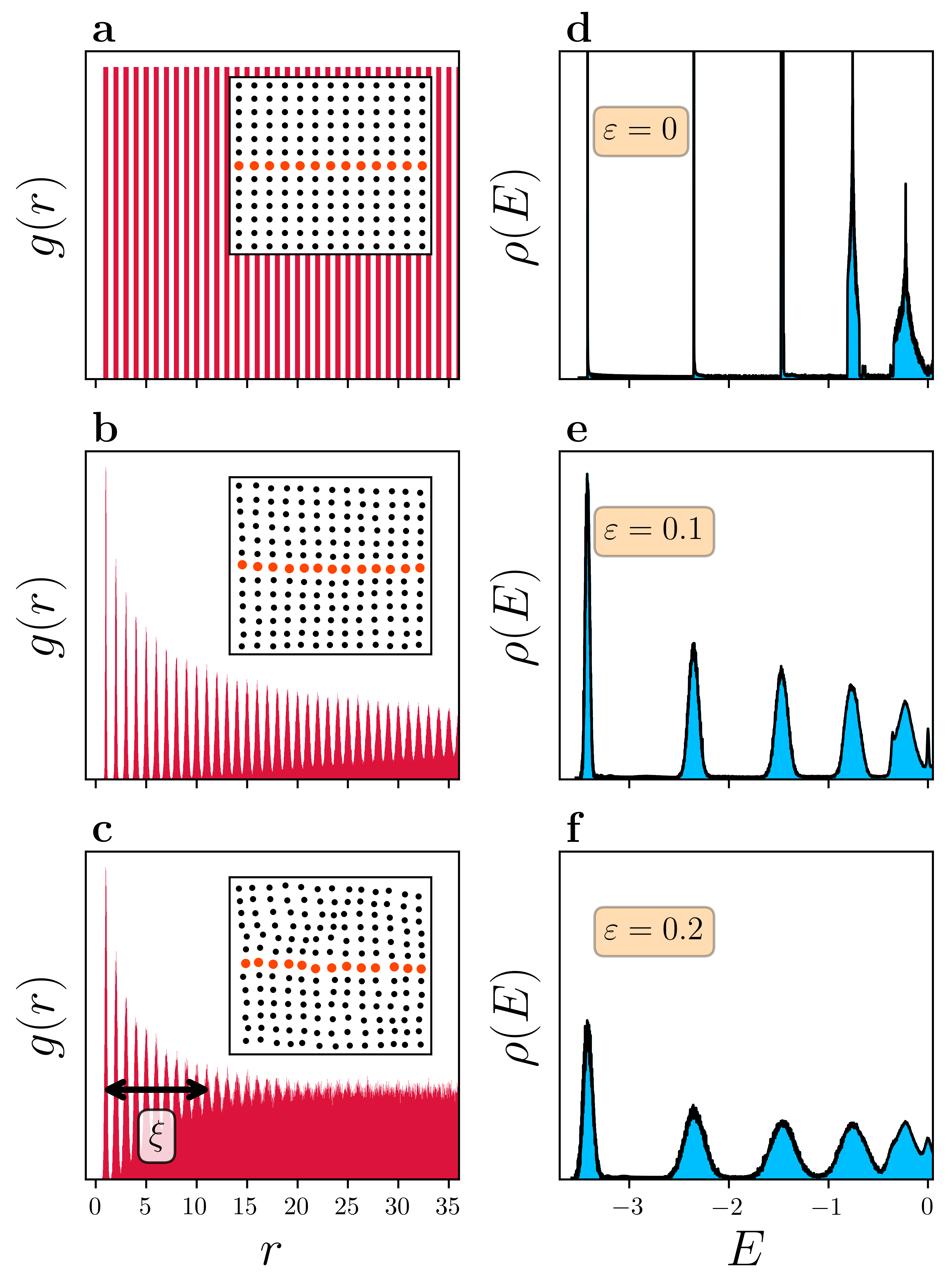}
\caption{\textbf{a}-\textbf{c}: Averaged density distribution function $g(r)=\sum_i \delta(r-r_i)$ of lattice sites along a horizontal cut in the studied lattices for three different values of the structural disorder parameter $\varepsilon$. The insets show an example of a small lattice configuration, and the red points indicate the direction for which the distribution was obtained. In the crystalline case $\varepsilon=0$, the distances are all integer multiples of the lattice constant, and the distribution remains uniform for all $r$. For non-zero $\varepsilon$, the long-range correlations decay after a length scale $\xi$, indicated by the black arrow for the $\varepsilon=0.2$ case.
\textbf{d}-\textbf{f}: Density of states $\rho(E)$ for magnetic flux $\phi=1/10$ per particle and $L=300$, for the same $\varepsilon$ as in the left panel.}
\label{fig1}
\end{figure}

In this work we study the fate of the Landau levels and quantum Hall effect on non-crystalline 2d lattices without long-range spatial order. On crystalline lattices, the Landau levels emerge in the low-field regime where the magnetic flux through a primitive cell is small compared to a flux quantum $\phi_0=h/e$ \cite{PhysRevB.14.2239}. These states, as their continuum counterparts, disperse linearly as a function of the magnetic field, are (nearly) flat and carry a unit Chern number~\cite{PhysRevLett.49.405}. We introduce a family of random lattices (depicted in the insets in Figs.~\ref{fig1} \textbf{a}-\textbf{c} which can be tuned continuously from a crystalline lattice, through a paracrystalline state~\cite{hindeleh1988paracrystals, hosemann1995structure}, to an amorphous phase~\cite{zallen2008physics}. While these lattices do not exhibit spatial long-range order, the density distribution function of a paracrystal exhibits residual periodic medium-range correlations~\cite{elliott1991medium, PhysRevLett.86.5514, cheng2011atomic}.
We establish that even when the scale of spatial correlations approaches the order of the average lattice constant i) the spectrum can still support well-resolved Landau-like bands shown in Figs.~\ref{fig1} \textbf{d}-\textbf{f}; ii) these bands disperse linearly in the magnetic field at low flux and support unit Chern numbers; iii) the system supports a quantized conductivity and topological edge modes. These facts suggest that also other phenomena arising from the Landau levels, typically associated with band structures, can survive even on random lattices without long-range spatial order.
\begin{figure}
\includegraphics[width=0.95\columnwidth]{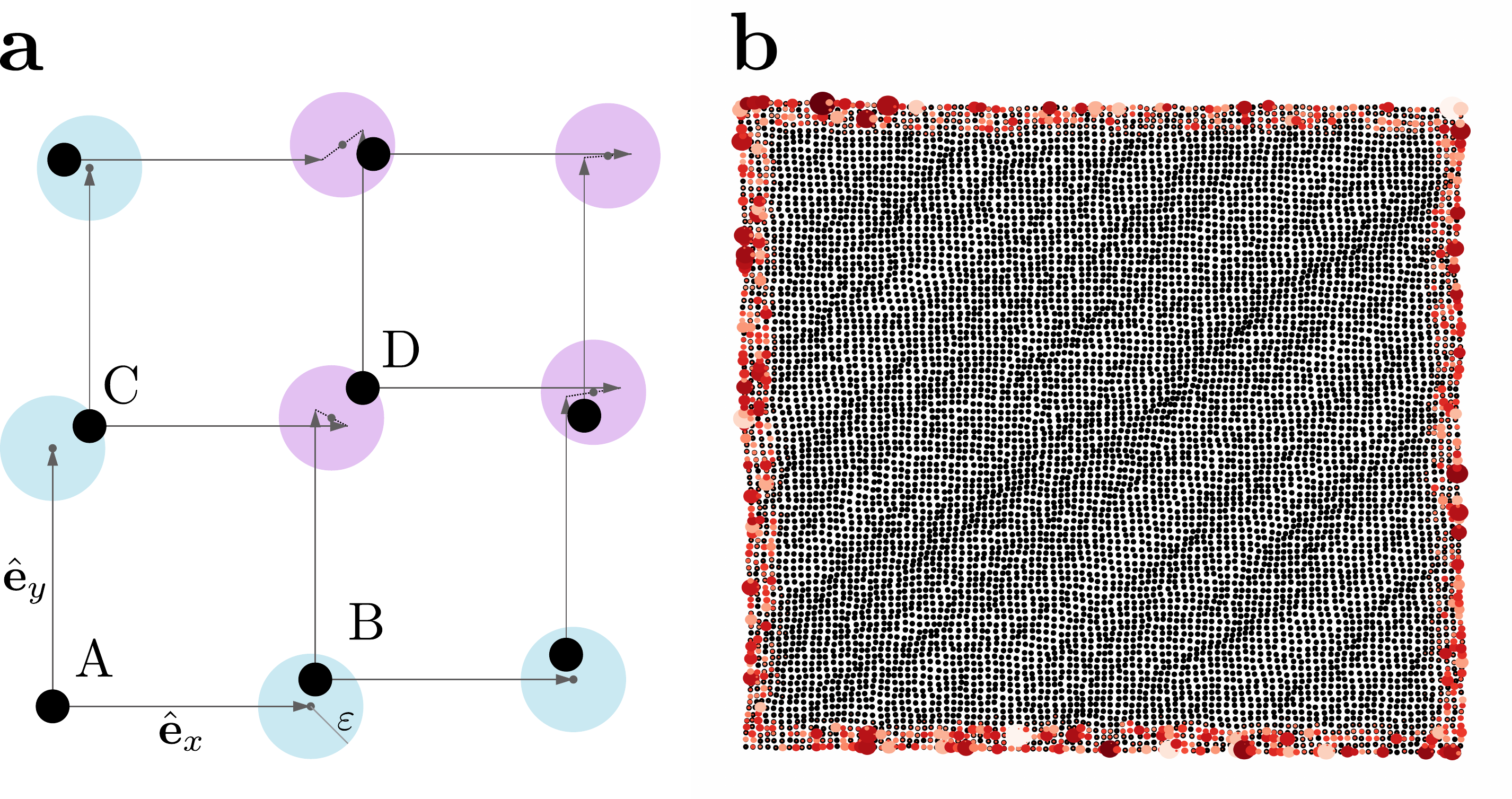}
\caption{\textbf{a}: Generation of lattices with tunable short- to medium-range spatial correlations.
\textbf{b}: Example of a lattice with $\varepsilon=0.2$ and linear size $L=100$.   The circles on the boundary layer represent the local DOS for flux $\phi=1/10$ and energy $E=-1.9$, revealing the topological edge modes for the Chern number $\mathcal C=2$ state. Size and colors of circles represent the magnitude of local DOS at each site.}
\label{fig2}
\end{figure}

\emph{Model--} The studied non-crystalline lattice geometries are characterized by short range square-lattice correlations, which decay over a characteristic correlation length $\xi$. 
The correlation length is controlled by the parameter $\varepsilon$, which characterizes a family of random lattices with the same statistical properties. In Figs.~\ref{fig1}~\textbf{a}-\textbf{c} we have plotted representatives of random lattices with different $\varepsilon$. The case $\varepsilon=0$ corresponds to a pristine square lattice with infinite correlation length, while $\varepsilon=0.1$ and $\varepsilon=0.2$ lead to random lattices with $\xi/a_0\sim 30$ and $\xi/a_0\sim 10$, respectively, in the horizontal direction as indicated in Fig.~\ref{fig1}. Setting the lattice constant $a_0$ to 1, the generation of the lattices and the role of $\varepsilon$ is illustrated in Fig.~\ref{fig2}~\textbf{a}. Starting from the origin point A, the adjacent site B is obtained by first translating A by $\hat{\bs{e}}_x$, drawing an $\varepsilon$-circle (blue) about that point, and randomly drawing a point B inside that circle, from a uniform distribution. The upper neighbour C is obtained analogously but translating A by vector $\hat{\bs{e}}_y$ instead. The diagonal neighbour D is obtained by translating C by $\hat{\bs{e}}_x$ and B by $\hat{\bs{e}}_y$, forming an $\varepsilon$-circle (purple) centered at the midpoint of the two translated points, and randomly placing D inside the circle. Repeating this process, one can generate arbitrarily large random lattices, one realization of which is illustrated in Fig.~\ref{fig2} \textbf{b}. The lattices generated by this prescription have the property that the local environment around any chosen interior point appears a perturbed square lattice. However, the lattice points more than distance $\xi$ away no longer appear at their square lattice positions with respect to a reference point, making it impossible to uniquely identify them with the square lattice sites. Thus, despite the short-range correlations, all geometries with $\varepsilon>0$ have no long-range correlation and cannot be regarded as perturbed crystalline lattices. Lattices with finite-range crystalline density correlations that are washed away for long distances, as seen in Fig.~\ref{fig1}, are referred to as paracrystals~\cite{hindeleh1988paracrystals, hosemann1995structure}. The studied lattices are homogeneous with statistically indistinguishable interior points; however, they are \textit{not} statistically isotropic, as indicated by the large-scale diagonal patterns, visible in Fig.~\ref{fig2} \textbf{b}, which breaks the four-fold rotation symmetry.

\begin{figure*}
\includegraphics[width=.95\linewidth]{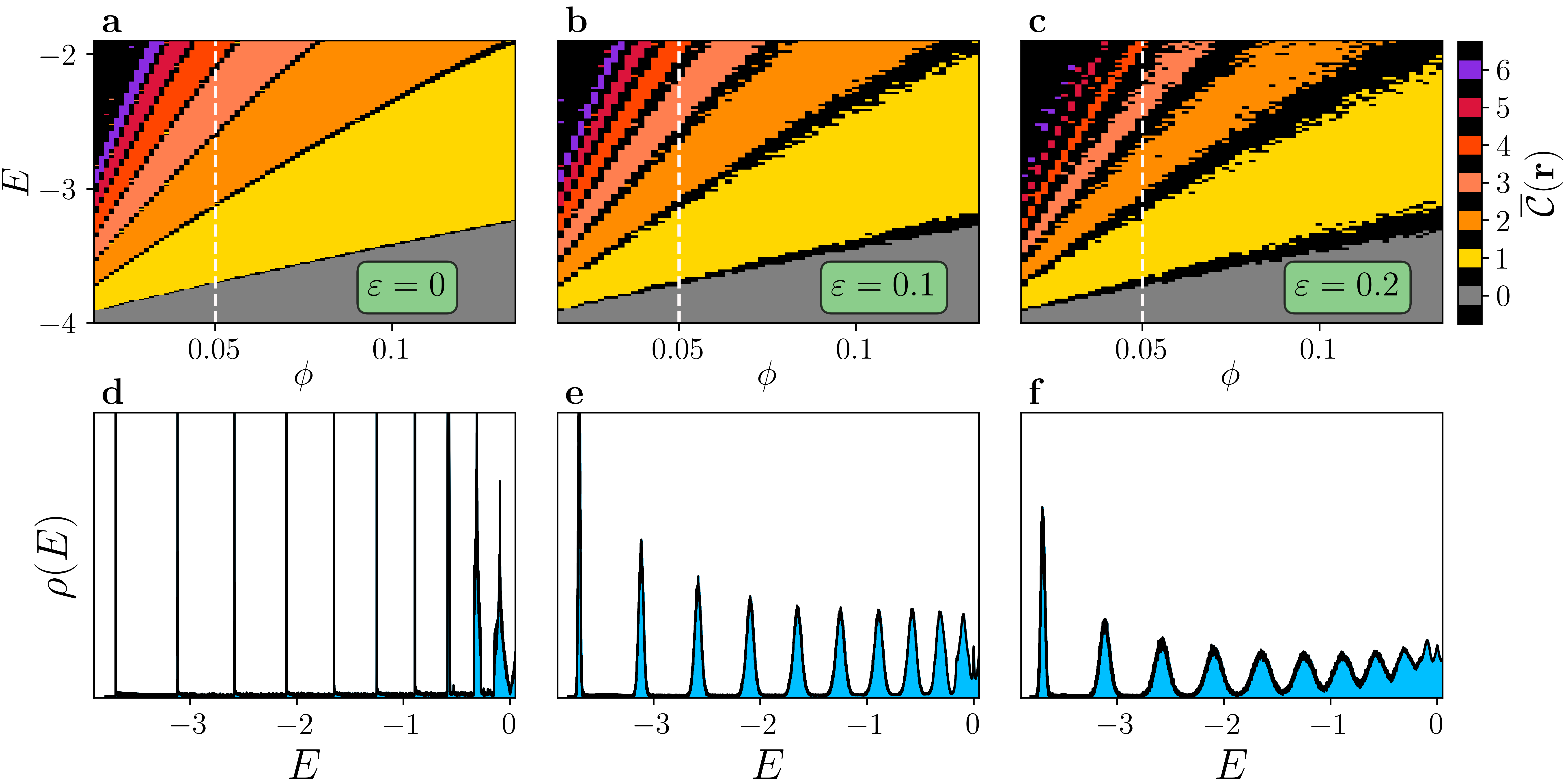}
\caption{\textbf{a}-\textbf{c}: Phase diagram for a $70\times 70$ system as a function of the structural disorder parameter $\varepsilon$, obtained from the averaged local Chern marker. For \textbf{b} and \textbf{c}, $\mathcal{\overline C}(\mathbf{r})$ is evaluated at every parameter pair for 200 different lattice configurations, and 100 randomly chosen points in the bulk of each lattice configuration. Colors are assigned for quantized $\mathcal{\overline C}(\mathbf{r})$, meaning values within 3\% of the indicated integers; higher Chern numbers are omitted, as they are barely discernable.
\textbf{d}-\textbf{f}: Density of states $\rho(E)$ for $L=500$ along the vertical cut $\phi=1/20$ indicated in white in the panel above.}
\label{fig3}
\end{figure*}

We now define a tight-binding model which describes hopping in a perpendicular magnetic field, generalizing the much-studied Hofstadter model~\cite{PhysRevB.14.2239} to the random lattices discussed above. We would like the coordination number to remain fixed for all studied geometries. Indeed, for sufficiently small $\varepsilon$, the four closest points of any interior point are the four nearest square lattice neighbours with perturbed distances. Thus, we can define the Hamiltonian as
\begin{equation}
    \mc{H}=t\sum_{<ij>} e^{-i\Phi_{ij}} \,\hat{c}^{\dagger}_i\, \hat{c}_j,
\end{equation}
where $t$ is the hopping parameter, $\hat{c}^{\dagger}_i$ ($\hat{c}_i$) is the creation (annihilation) operator at site $i$, and the hoppings connect each site to its four closest points, as the sum is over all neighboring pairs.
The Peierls phases $\Phi_{ij}=e/\hbar\int_i^{j}\mathbf{A} \cdot d\mathbf{l}$ accommodate the magnetic field which arises from the vector potential $\mathbf{A}$. The magnetic field is assumed to be spatially homogeneous, so the vector potential can be written as $\mathbf{A}(\vec{r})=\frac{B}{2} (x\hat{\bs{e}}_y-y\hat{\bs{e}}_x)$ in the symmetric gauge and $\mathbf{A}(\vec{r})=B x\hat{\bs{e}}_y$ in the Landau gauge. We have employed different gauges to confirm that the results for various observables presented below are gauge-invariant. For the studied random lattices, the average distance between the closest lattice points is fixed by a square lattice constant $a_0$. Also, the average magnetic flux per site remains fixed to its square lattice value $\phi=Ba_0^2/\phi_0$ for lattices corresponding to different $\varepsilon$, which makes it an appropriate quantity to parametrize the properties of the system.
One could also introduce a variable hopping strength; however, due to small variations in bond lengths, we elect to keep the number of free parameters as low as possible and focus on the crucial magnetic-field-dependent modifications of complex phases.

\emph{Landau levels and Quantum Hall effect--} We are now ready to explore the physical properties of the model. We concentrate on the low flux regime $\phi\ll 1$, which is relevant for real solids and makes connection to the Landau levels in the continuum.

We start by evaluating the topological phase diagram, determined by the Chern numbers, and the density of states (DOS) defined as
\begin{equation} \label{eq:dos}
\rho(E)=L^{-2} \sum_n\delta(E-E_n).
\end{equation}
Additionally, an efficient method to extract the phase diagram in random systems is to evaluate the local Chern marker~\cite{PhysRevB.84.241106, caio2019topological,  PhysRevResearch.2.013229} and average it over many lattice sites, and over different random configurations to obtain the Chern number. The local Chern marker is defined by
\begin{equation}
    \mc C (\vec r) = -2\pi i\int \mathrm d \vec r' \left[\mc X(\vec r, \vec r') \mc Y(\vec r',\vec r) - \mc Y(\vec r, \vec r') \mc X(\vec r',\vec r)\right],
\end{equation}
where $\mc X(\vec r, \vec r') = \int \mathrm d \vec r'' P(\vec r,\vec r'') x'' P(\vec r'',\vec r')$ is the projected $x$ coordinate, and $\mc Y(\vec r, \vec r')$ is similarly defined projected $y$ coordinate. Here $P(\vec r,\vec r')$ denotes the matrix elements of the projection operator to the filled energy bands. In practice, to avoid diagonalization of the full system, we calculate the DOS and the Chern marker by employing the Kernel Polynomial Method (KPM)~\cite{RevModPhys.78.275}. While the calculation of the DOS using the KPM is well-documented~\cite{RevModPhys.78.275}, the derivation of the kernel polynomial formula for the Chern marker and the details of the configuration averaging  are presented in the SI \footnote{see Supplementary Information for the kernel polynomial implementation of the Chern marker}.
\begin{figure*}
\includegraphics[width=0.9\linewidth]{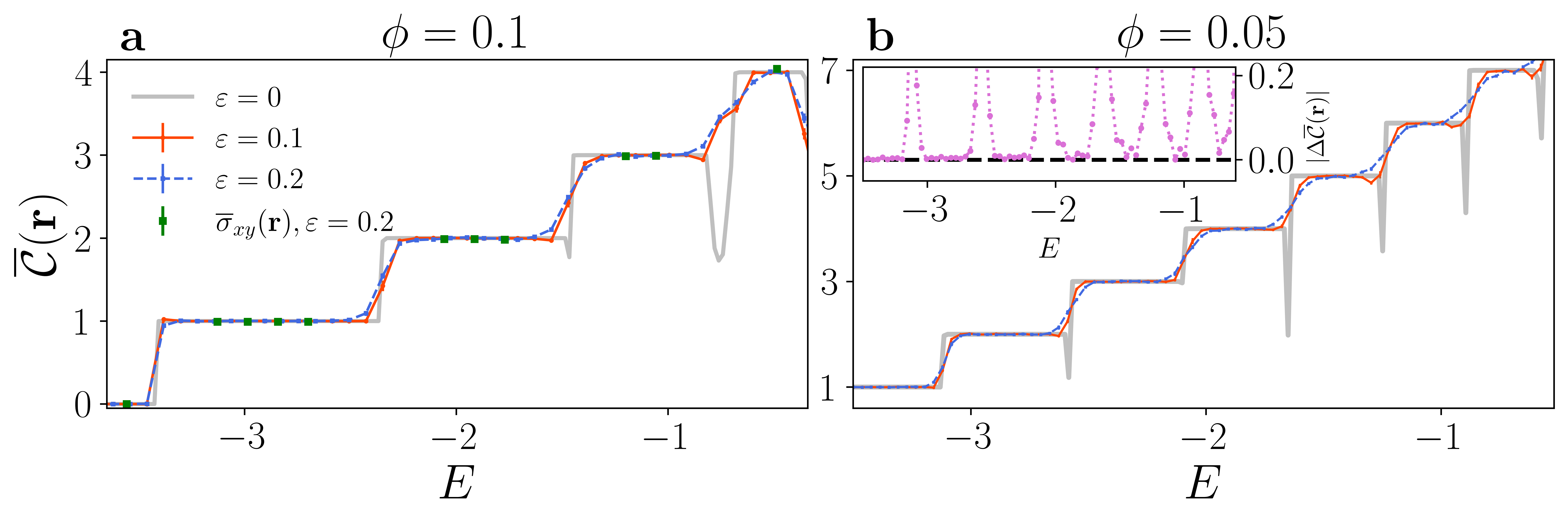}
\caption{Averaged local Chern marker $\overline{\mathcal C}(\mathbf{r})$ as function of energy for different $\varepsilon$ for fluxes  $\phi=1/10$ (\textbf{a}) and $\phi=1/20$ (\textbf{b}). In \textbf{a} we plot the Hall conductivity $\overline{\sigma}_{xy}$ evaluated at selected energies and averaged over random lattice positions deep in the bulk for the $\varepsilon=0.2$ case.
The inset in \textbf{b} illustrates the absolute deviation in the averaged local Chern marker values $|\Delta \overline{\mathcal C}(\mathbf{r})|$ between $\varepsilon=0.2$ and the crystalline case. Here the linear size of the system is $L=200$, and the local Chern values are averaged over $n=20$ random configurations and $150$ random lattice positions for each configuration. Error bars are not visible in this scale.}
\label{fig4}
\end{figure*}


%
In Figs.~\ref{fig3} \textbf{a}-\textbf{c} we have illustrated how the topological phase diagram evolves as the spatial correlation length is decreased. The corresponding DOS for each case is presented in Figs.~\ref{fig3} \textbf{d}-\textbf{f}. The DOS and the phase diagram are symmetric with respect to $E=0$, so we plot them only for negative energies. In the crystalline limit $\varepsilon=0$, we recover the almost flat Landau bands, which disperse linearly in the low flux regime as expected. These states also carry a unit Chern number, as indicated by the unit jumps in the Chern marker at energies corresponding to the Landau bands. For finite $\varepsilon$, the DOS corresponding to individual Landau bands becomes significantly broadened, however, the main features of the phase diagram survive even in the presence of strong randomness $\varepsilon= 0.2$. The relevant length scales for the problem are the average lattice constant $a_0$, the magnetic length $L_{B}=1/\sqrt{2\pi \phi}$ (the characteristic length of the Landau orbits), and the spatial correlation length $\xi$ of the lattice. For the parameters corresponding to Figs.~\ref{fig3} \textbf{c} and \textbf{f}, they satisfy $a_0\lesssim L_B\sim 0.1 \xi$, indicating that the Landau level picture and the quantum Hall effect persists even when the correlation length approaches the microscopic length scales of the problem. The broadening caused by the random lattice geometry on the Landau bands appears qualitatively similar to that caused by disorder on crystalline lattices and in continuum~\cite{huckestein}. However, this observation is far from self-evident, since disorder can be regarded as a mere perturbation to the underlying structure which always retains an approximate translation invariance. In contrast, the finite correlation length in non-crystalline lattices cannot be regarded as a perturbative effect and there exists no systematic approach in which a perfect crystal would serve as a starting point for the analysis.
Furthermore, the studied model is fundamentally different from those obtained by perturbing the sites of a crystalline system independently around their regular positions. Such lattices support arbitrary long-range spatial order with $\xi = \infty$, and the density distribution function associated with them resembles Fig.~\ref{fig1}~\textbf{a}, except with broadened delta peaks.

A non-zero Chern number gives rise to the quantum Hall effect, indicating the existence of a quantized Hall conductivity in the system. As shown in Fig.~\ref{fig4}~\textbf{a}, this is confirmed by a direct evaluation of the Hall conductivity using a method developed in Ref.~\cite{PhysRevLett.114.116602} based on a real-space Kubo-Bastin formalism~\cite{PhysRevB.91.165117,bastin1971quantum, groth2014kwant}. The evolution of the quantized plateaus as a function of $\varepsilon$, illustrated in Fig.~\ref{fig4} \textbf{b}, indicates that the overlapping higher Landau bands will be the first affected by randomness. When the Landau level broadening increases, the plateaus shrink, and when the broadening becomes comparable to the level separation they eventually disappear~\cite{PhysRevLett.78.318}. In addition to the quantized Hall conductance, the non-trivial Chern numbers also give rise to topologically protected edge modes at energies that fall in the  mobility gap between the Landau levels~\cite{PhysRevB.25.2185}. These edge modes are revealed by the local DOS as seen in Fig.~\ref{fig2} \textbf{b}. The intensity of the local DOS varies randomly along the edge, reflecting the irregular geometry of the lattice, vanishing rapidly beyond the surface layer.

\emph{Discussion--}  While the existence of amorphous topological states has been established in various systems recently~\cite{PhysRevLett.118.236402,PhysRevB.96.121405,mitchell2018amorphous,poyhonen2018amorphous,bourne2018non,PhysRevLett.123.076401,zhou2020photonic,marsal2020topological,PhysRevResearch.2.043301, 10.21468/SciPostPhys.11.2.02, PhysRevLett.128.056401, ivakifractal, PhysRevLett.130.026202}, the present work is the first study of the Landau levels as a function of spatial correlations. It should be emphasized that our findings of the non-crystalline Landau bands have significant implications beyond the quantum Hall effect. Since the birth  of modern solid state physics, it has been known that many thermodynamic and transport quantities are determined by the DOS at the Fermi level, resulting in quantum oscillations when the Landau levels are pushed through the Fermi level by varying the magnetic field~\cite{solyom,girvin}. Qualitatively similar oscillations persist in the studied systems, as indicated by the Landau peaks in the DOS that also disperse as a function of the magnetic field. The Landau level formation and the resulting quantum oscillations are typically considered as a consequence of a band structure, so it is remarkable that the quantum oscillations may survive in systems where the notion of a band structure does not apply. Approximately flat bands which carry finite Chern numbers are also a prerequisite for more complex interaction-induced phases, suggesting that the lack of a long-range spatial order does not pose a fundamental obstacle for formation of fractional quantum Hall states~\cite{ezawa2013quantum,girvin}. 

While in this work the focus was on 2d systems, the phenomena resulting from the existence of Landau levels have a wide variety of consequences in 3d systems~\cite{solyom,gooth2022quantum}. Our results indicate that, at least qualitatively, many hallmark features of 3d electrons in magnetic fields can be expected to survive in non-crystalline 3d systems without long-range order. However, a systematic study of these features in 3d quickly becomes computationally challenging.

\emph{Summary--} In this work we studied the evolution of Landau levels and topology as a function of randomness in non-crystalline 2d lattices without long-range order. We discovered that, even when the spatial correlations decay over microscopic length scales, the Landau levels persist as broadened but well-defined features. Moreover, we showed that the magnetic field dependence and the topological nature of the Landau bands remain intact, giving rise to the quantum Hall effect with quantized conductance and topological edge modes. The fact that the lowest Landau bands can be resolved, despite the lack of long range order, imply that the fundamental quantum effects that have been traditionally considered as properties of crystalline solids in the magnetic field, can also be observed in non-crystalline systems.

\emph{Acknowledgement--} This work was supported by the Academy of Finland through the project number 331094. 


\bibliography{ref.bib}

\clearpage
\newpage

\widetext
\begin{center}
\Large SUPPLEMENTAL INFORMATION to ``Quantum Hall effect and Landau levels without spatial long-range correlations''
\end{center}

\section{Local Chern marker from kernel polynomial expansion}
The local Chern marker can be defined by the equation
\begin{equation}
	\mc C (\vec r) = -2\pi i\int \mathrm d \vec r' \left[\mc X(\vec r, \vec r') \mc Y(\vec r',\vec r) - \mc Y(\vec r, \vec r') \mc X(\vec r',\vec r)\right] \equiv \mc C_{xy}(\vec r) - C_{yx}(\vec r),
\end{equation}
where
\begin{equation}
	\mc X(\vec r, \vec r') = \int \mathrm d \vec r'' P(\vec r,\vec r'') x'' P(\vec r'',\vec r'),
\end{equation}
and analogously for $\mc Y(\vec r, \vec r')$, in terms of the projection operator to occupied bands
\begin{equation}
	P(\vec r, \vec r') = \int_{E_\text{min}}^{E_\text{max}} \mathrm d E \theta(E_F-E)\sum_n \delta(E - E_n)\braket{\vec r}{n}\braket{n}{\vec r'},
\end{equation}
with $n$ labeling the eigenstates of the Hamiltonian. In order to implement KPM we rescale energies to lie in the interval $(-1,1)$:
\begin{align}
	\mc H &= (H - b)/a\\
	a &= (E_\text{max} - E_\text{min})/(2 - \eta)\\
	b &= \frac{1}{2}(E_\text{max} + E_\text{min}).
\end{align}

Together with the substitution $\epsilon = (E-b)/a$, we get
\begin{equation}
	P(\vec r, \vec r') = \int_{-1}^{1} \mathrm d \epsilon \theta(\epsilon_F-\epsilon)\sum_n \delta(\epsilon - 
	\epsilon_n)\braket{\vec r}{n}\braket{n}{\vec r'},
\end{equation}
and hence, after performing the radial integrals, 
\begin{equation}
	C_{xy} = \int_{\lbrace -1,1\rbrace^3} \mathrm d^3 \epsilon \sum_{nmk} \braket{\vec r}{n} \bra{n} \vec{\hat x} \ket m \bra m  \vec{\hat y} \ket k \braket{k}{\vec r} \delta(\epsilon_1-\epsilon_n)\delta(\epsilon_2-\epsilon_m)\delta(\epsilon_3-\epsilon_k) \theta(\epsilon_F-\epsilon_1)\theta(\epsilon_F-\epsilon_2)\theta(\epsilon_F-\epsilon_3).
\end{equation}

Note that the integration here is over the numbered epsilons, while the ones labelled by letters are instead summed over. We now approximate this result by replacing the quantities
\begin{equation}
 f(\epsilon_1) = \sum_{n} \ket{n} \bra{n}  \delta(\epsilon_1-\epsilon_n),
\end{equation}
with their truncated Chebyshev expansions, i.e.
\begin{equation}
	f(\epsilon_1) \approx f_\text{KPM}(\epsilon_1) = \sum_{n=0}^{N-1} \frac{2\mu_{n}h_{n} g_n T_n(\epsilon_1)}{\pi(1 + \delta_{n,0})\sqrt{1-\epsilon_1^2}} .
\end{equation}

Here $\mu_{n}$ are the moments
\begin{align}
	\mu_{n} &= \int_{-1}^1 \mathrm d \epsilon f(\epsilon_1) T_n(\epsilon_1) = T_n(\mc H),
\end{align}
and $g_n$ represent the kernel terms improving the accuracy of the truncated sum. Inserting this into the expression for $C_{xy}$, we get
\begin{align}
    C_{xy} &= \int_{\lbrace -1,1\rbrace^3} \mathrm d^3 \epsilon \sum_{nmk} \bra{\vec r} f_\text{KPM}(\epsilon_1) \vec{\hat x} f_\text{KPM}(\epsilon_2)\vec{\hat y} f_\text{KPM}(\epsilon_3) \ket{\vec r}  \theta(\epsilon_F-\epsilon_1)\theta(\epsilon_F-\epsilon_2)\theta(\epsilon_F-\epsilon_3)\notag\\
	&= \bra{\vec r} \mc T(\mc H)\vec{\hat x}  \mc T(\mc H)\vec{\hat y}  \mc T(\mc H)\ket{\vec r} ,
\end{align}
where
\begin{equation}
 \mc T(\mc H) = \sum_n \frac{2g_nT_n(\mc H)}{(1 + \delta_{n,0})} \int_{-1}^1 \mathrm d \epsilon \frac{T_n(\epsilon) \theta(\epsilon_F - \epsilon)}{\pi \sqrt{1-\epsilon^2}}.
\end{equation}

Approximating the integral and subsequently making use of the discrete Fourier transform (see Ref.~\onlinecite{RevModPhys.78.275}), we obtain
\begin{equation}
	\mc T(\mc H) \approx \sum_n \frac{2g_nT_n(\mc H)}{(1 + \delta_{n,0})} \frac{1}{M} \sum_{s=0}^{M-1}\cos \left(\frac{n \pi(s + \frac{1}{2})}{M}\right) \theta\left[\epsilon_F-\cos\left(\frac{\pi(s + \frac{1}{2})}{M}\right)\right] = \sum_n \frac{2g_nT_n(\mc H)}{M(1 + \delta_{n,0})} T_n(\mc H)  S(n),
\end{equation}
where we assume $M > N$.

We note that, as a function of $s$, the argument inside the step function increases monotonically; as $|\epsilon_F| < 1$, the overall effect of the step function is hence to exclude terms from the beginning of the sum. Let us denote $\Gamma \equiv \ceil{\frac{M}{\pi}\arccos \epsilon_F - \frac{1}{2}}$.
Then,
\begin{equation}
	S(n) = \sum_{s=\Gamma}^{M-1}\cos \left(\frac{n \pi(s + \frac{1}{2})}{M}\right) = \text{Re}  \left[ \sum_{s=\Gamma}^{M-1} e^{\frac{n \pi(s + \frac{1}{2})}{M}}\right] = \text{Re} \left[e^{\frac{in\pi}{2M}}e^{\frac{in\pi\Gamma}{M}}\sum_{j=0}^{M-\Gamma-1}e^{\frac{n \pi j}{M}}\right].
\end{equation}

This yields
\begin{equation}
\begin{cases}
    \hspace{0.2cm} S(n) = M - \Gamma, \hspace{1cm} &n = 0 \\
    \hspace{0.2cm} S(n) = -\frac{1}{2}\frac{\sin \frac{n \pi \Gamma}{M}}{\sin \frac{n\pi}{2M}}, &n > 0 .
\end{cases}
\end{equation}

Absorbing a minus sign into $\mc T$, we finally get
\begin{equation}
	\mc C (\vec r) = 4\pi \text{Im}  \bra{\vec r} \mc T(\mc H)\vec{\hat x}  \mc T(\mc H)\vec{\hat y}  \mc T(\mc H)\ket{\vec r}  
\end{equation}
where, using $T_0(x) = 1$ and $g_0 = 1$,
\begin{equation}
	\mc T(\mc H) \approx -1 + \frac{\Gamma}{M} + \sum_{n=1}^{N-1} \frac{g_n}{M} \frac{\sin \frac{n \pi \Gamma}{M}}{\sin \frac{n\pi}{2M}} T_n(\mc H).
\end{equation}

\section{Calculation details}
\begin{figure*}[h]
\includegraphics[width=0.95\linewidth]{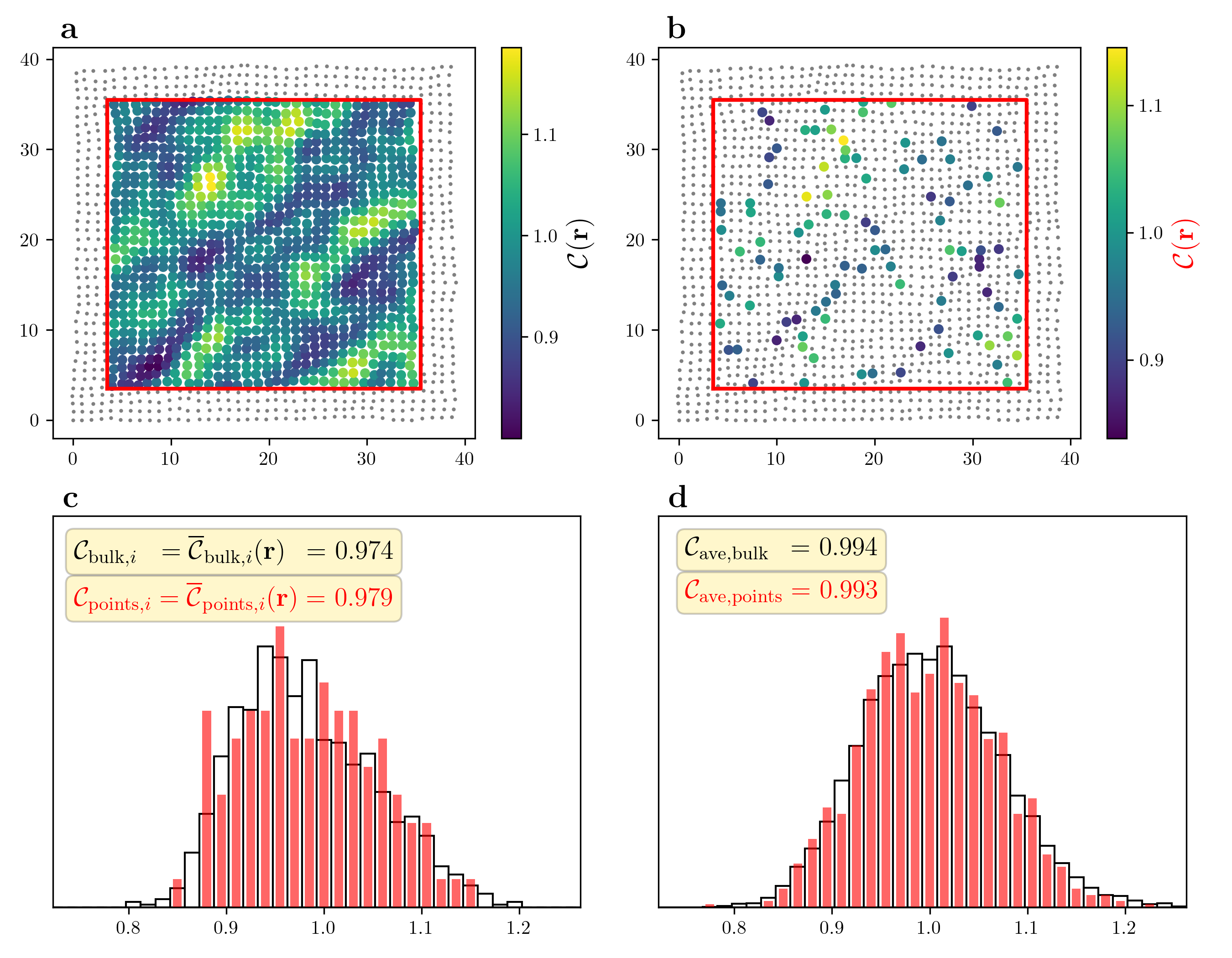}
\caption{\textbf{a}: Sampling of the local Chern marker for all bulk points in a given  configuration. \textbf{b}: Local Chern marker shown for a subset of $N=100$ points. The parameters used are $\phi=0.1$ and $E=-3$ with the structural disorder parameter $\varepsilon=0.2$. \textbf{c}: The normalized full distribution of local marker values, shown in white, as well as the partial distribution of random points in light red, for the single configuration used in \textbf{a} and \textbf{b}. \textbf{d}: Same as \textbf{c}, but for a set of 10 randomly generated configurations. The distributions match very well and yield the same mean value for the Chern number with a high accuracy.}
\label{fig:app_distribution}
\end{figure*}

For quantities such as the Chern marker and conductivity tensor which are evaluated locally, only a finite number of lattice points in the bulk are sufficient to fully describe topological features of the system.
The local markers vary significantly throughout the lattices with open boundaries, but can form small regions where the changes in the values are almost continuous. This means that choosing to average many points next to each other may be a biased sample of the distribution, depending on the configuration.
Therefore, rather than e.g.\ choosing a large region near the center, we randomly choose points within the bulk. This approach has the benefit of sampling the entirety of the lattice, excluding only a finite border close to the edges. We typically choose $N \sim 100$ points randomly from the bulk where we evaluate the local marker, yielding an appropriate average for the quantity at hand. In other words, it is an unbiased sampling of the full distribution of, say, local Chern marker values $\mathcal C(\mathbf{r})$, and thus we can obtain a proper average value with less computational cost, especially when we scale up the system size. We repeat this sampling for a number of configurations, and the distributions converge together remarkably quickly, as shown in Fig.~\ref{fig:app_distribution}.
This observation provides a powerful and efficient tool to study larger systems. The number of random samples needed to describe the correct distributions and the corresponding mean values remains roughly constant, provided $N$ is not too small.

Moreover, for an accurate description within KPM, in our calculations we set the number of moments $N_m$ used in the expansion to $N_m\approx L^2/3$ for systems which host $L^2$ electronic states. Additionally, in studying DOS, the accuracy in calculating stochastic traces depends on the product $N_r\,N_c$, where $N_r$ is the number of random phase vectors and $N_c$ is the number of independent random lattice configurations. In our numerical simulations $N_rN_c \sim 10^3$. We also studied the averaged Hall conductivity $\sigma_{xy}(\mathbf{r},E)$ of the bulk, similarly to the calculation of Chern markers by random sampling of lattice points away from the boundaries, and found that it matches well with the predicted Chern number. This hints toward the equivalence of the \textit{local} bulk conductivity, which is a physical observable, and the local Chern marker.


\end{document}